\documentclass[11pt,preprint,superscriptaddress,aps,showkeys,nofootinbib]{revtex4}
\usepackage{amssymb,amsmath,amsfonts}
\usepackage{graphicx}
\usepackage{graphics}
\usepackage{eepic,epsfig}
\usepackage{indentfirst}
\usepackage[brazilian]{babel}
\usepackage[utf8]{inputenc}
\usepackage[T1]{fontenc}
\usepackage{bm} 

1.60cm \makeatletter \@addtoreset{equation}{section}

\makeatother

\begin{document}
\title{Fermionic Casimir effect in Horava-Lifshitz theories}
\author{D\^eivid R. da Silva}
\affiliation{Departamento de F\'{\i}sica, Universidade Federal da Para\'{\i}ba\\
	Caixa Postal 5008, 58051-970, Jo\~ao Pessoa, Para\'{\i}ba, Brazil}
\email{drds.sensei@gmail.com, messiasdebritocruz@gmail.com, emello@fisica.ufpb.br}
\author{M. B. Cruz}
\affiliation{Departamento de F\'{\i}sica, Universidade Federal da Para\'{\i}ba\\
	Caixa Postal 5008, 58051-970, Jo\~ao Pessoa, Para\'{\i}ba, Brazil}
\email{drds.sensei@gmail.com, emello@fisica.ufpb.br, messiasdebritocruz@gmail.com}
\author{E. R. Bezerra de Mello}
\affiliation{Departamento de F\'{\i}sica, Universidade Federal da Para\'{\i}ba\\
	Caixa Postal 5008, 58051-970, Jo\~ao Pessoa, Para\'{\i}ba, Brazil}
\email{drds.sensei@gmail.com, emello@fisica.ufpb.br, messiasdebritocruz@gmail.com}


\begin{abstract}
In this paper, we analyze the fermionic Casimir effects associated with a massless quantum field in the context of Lorentz symmetry violation approach based on Horava-Lifshitz methodology. In order to obtain these observables, we impose the standard MIT bag boundary condition on the fields on two large and parallel plates. Our main objectives are to investigate how the Casimir energy and pressure depend on the parameter associated with the breaking of Lorentz symmetry. 
\end{abstract}

\maketitle


\section{Introduction}
In the Quantum Field Theory (QFT), the Hamiltonian operator associated to quantum free fields can be considered as the sum of an infinite set of quantum harmonic oscillators. Under this point of view, the vacuum state can be understood as the state where all these quantum oscillators are in their ground states. Consequently, the vacuum energy,  being the sum of the energies of the ground states of these oscillators, is infinite. It is well known that QFT presents several examples in which it is exhibited that the vacuum state plays a fundamental role not only in the physics of microscopic phenomena \cite{Bjorken:1965zz} but also the physics of macroscopic phenomena. One of these most important phenomena is the Casimir effect \cite{Casimir:1948dh}.

Originally, the Casimir effect was proposed in $1948$, under a theoretical point of view, by considering the vacuum state associated with electromagnetic fields confined between two large, parallel and conductor plates. Due to the boundary condition imposed on the quantum fields on the two plates, only vacuum excitation with specific wavelengths are allowed between them. In this sense, the Casimir energy can be obtained by subtraction from the energy associated with these quantum fluctuations, the energy of the vacuum state in the absence of boundaries. Adopting this renormalization procedure, Casimir found an attractive force between the plate, which is given by:
\begin{equation}
F = - A \frac{\pi^2 \hbar c}{240 a^4}\   ,
\end{equation}
where $A=L^2$ is the area of plates and $a<<L$ is the distance between them.
This effect was experimentally confirmed in \cite{Sparnaay:1958wg}. In $90$s, experiments have confirmed the Casimir effect with a high degree of precision \cite{Lamoreaux:1996wh, Mohideen:1998iz,Harris:2000zz}.

Although only the Casimir effect associated with the electromagnetic field has been experimentally observed, in principle this effect also occurs for all relativistic quantum field, like scalar and fermionic, submitted to specific boundary conditions. In \cite{Johnson:1975zp} the authors have used the MIT bag model to confine quarks in hadrons admitting a spherical shape to them. The analysis of the Casimir effect associated with massless scalar and fermionic quantum fields confined between two large plates is presented in \cite{Bordag:2009zzd}. 

For a fermionic field with mass $m$, the Casimir energy associated with the field confined between two large parallel plates of area $A$, separeted by a distance $a$, can be expressed in terms of an integral representation \cite{Aram}, or by an infinity series expansion evolving derivative of the Whittaker function \cite{Elizalde}. For both representations approximated expressions can be provided. For large value of the dimensionless parameter $am$, the Casimir energy presents a exponential decay, $e^{-2am}$, and for $am<<1$, the leading term in the the Casimir force is a mass independent. In fact, up to the first order correction, the Casimir force reads,
\begin{eqnarray} 
\label{Force}
F= -A\frac{\big{(}7\pi^2-80 am \big{)}\hbar c}{960 a^4} \ .
\end{eqnarray}
By the above result we can see that, a possible experiment evolving fermionic fields would be preferably appreciable for neutrinos fields.

As we know, the standard QFT is based on the Lorentz symmetry invariance of the full theory. However, with the objective to construct a renormalized quantum field theory to gravity, Horava \cite{Horava:2009uw} has proposed a model, named Horava-Lifshitz (H-L) model, that presents an anisotropy between space and time. In this way, the Lorentz symmetry is broken in a strong manner. Although the problem of renormalizability of quantum gravity takes place at scale comparable with the Planck energy, the  gravitation beyond the General Relativity have shown that relic signature of these models can be present in Quantum Electrodynamics \cite{Liberati}. So, the space-time anisotropy in a given field theory model certainly modifies, at classical level the dispersion relation, and consequently the spectrum of the corresponding Hamiltonian operator. 

One of the most recent experiment that presents a very high degree of precision is the Casimir effect. So, this experiment is a good candidate to analyze a possible deviation caused by the consequence of Lorentz violation on the Casimir energy.  So, with the objective to investigate how the Casimir energy is affected by adopting a Lorentz symmetry violation in the context of H-L, in \cite{Ferrari:2010dj,Ulion:2015kjx} the authors have analyzed the Casimir energies and pressures, associated with massless scalar quantum fields confined between two parallel plates, admitting that the fields obey the Dirichlet, Newman and mixed boundary conditions on the plates. 

Besides the H-L approach, in $1989$ V. A. Kostelecky and S. Samuel \cite{Kostelecky:1988zi} described a mechanism in string theory that allows the violation of Lorentz symmetry at the Planck energy scale. In this mechanism, the Lorentz symmetry is spontaneously broken through the emergence of a non-vanishing vacuum expectation values of some vector and tensor components. In this way, a  preferential direction for space or time is implemented. In this context the analysis of scalar Casimir energy and pressure have been analyzed in \cite{Cruz:2017kfo} admitting specifics boundaries condition obeyed by the field on two parallel plates; moreover, the influence of the non-vanishing temperature on these observables has been considered in \cite{Cruz:2018bqt}. Finally, the analysis of Casimir energy and pressure associated with fermionic quantum fields obeying the MIT bag boundary condition on two parallel plates, was investigated in \cite{PhysRevD.99.085012}.

The main objective of this paper is to investigate the Casimir energy and pressure, associated with the massless fermionic quantum field in an H-L approach. In this way, we will consider that the fermionic field, solution of the modified Dirac equation, obeys the standard MIT bag boundary condition on two parallel plates of area $L^2$, separated by a distance $a$, being $a<<L$.

This paper is organized as follows: In Section \ref{secII} we consider the modified Dirac Lagrangian in the context of H-L approach. We present explicitly, for massless fields, its solutions compatible with the MIT bag boundary condition, taking into consideration, separately, that the parameter associated with the Lorentz symmetry breaking, named critical exponent, $\xi$, is an even or an odd number. As we will see, there appear a profound modification in the obtainment of the solutions for these two distinct situations.  In Section  \ref{sec:Casimir_effect} we calculate the Casimir energies and forces for the system under consideration. Specifically, we develop, separately, these calculations considering even and odd values for $\xi$. In Section \ref{sec:conclusions} we provide our most relevant remarks. Finally, in Appendix \ref{app}, we analyze the consequences of applying the standard MIT bag model on the current densities generated by the continuity equations of the modified Dirac equation. We argue that this condition prevents the current density obtained to cross the two flat boundaries.


\section{Dirac Equation in the Horava-Lifishitz Formalism}
\label{secII}
In this section, we want to analyze the modified Dirac equation in the context of H-L approach. Specifically, we want to obtain solutions for this equation, compatible with the standard MIT bag boundary condition imposed on the fields on two parallel plates of areas $L^2$, separately by a distance $a<<L$.

Following the model proposed in \cite{Farias:2011aa}, the Lagrangian density adopted to describe the dynamics of a massless fermionic field in the context of H-L methodology is given by: 
\begin{equation}
\label{eq:lagrangian_violation}
	\mathcal{L} = \bar{\psi} \left[ i\gamma^{0}\partial_t + i^{\xi} l^{\xi - 1} \left( \gamma^j \partial_j  \right)^{\xi} \right]\psi.
	\end{equation}
	
The exponent $\xi$, named the critical exponent, is associated with the  Lorentz symmetry violation. For $\xi=1$, we recover the usual Dirac Lagrangian. Moreover, the additional parameter $l$, with the dimension of the inverse of mass, has been introduced to provide the correct dimension for the Lagrangian. Here in this paper, we are adopting the Bjork-Drell notation \cite{Bjorken:1965zz} for the  $\gamma$ matrices:
	\begin{equation}
	\gamma^0 = \left( \begin{array}{cc}
	I & 0 \\ 
	0 & -I
	\end{array} \right) \quad \mathrm{and} \quad \gamma^j = \left( \begin{array}{cc}
	0 & \sigma^j \\ 
	- \sigma^j & 0
	\end{array}  \right)\cdot
	\end{equation}
	
\begin{figure}[hbtp]
	\centering
	\includegraphics[scale=0.3]{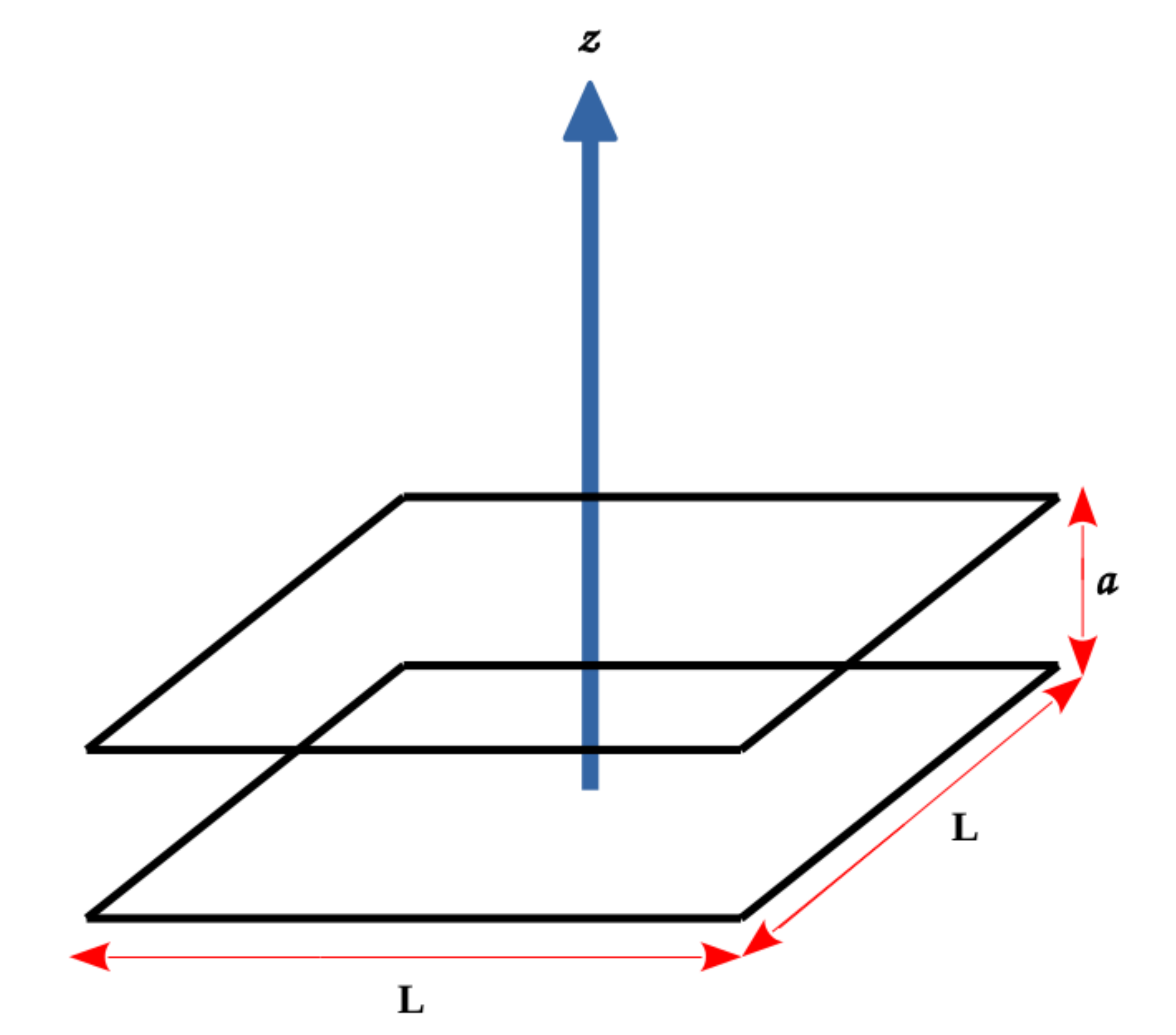}
	\caption{Two identical parallel plates separated by a distance $a$.}
	\label{fg:plates}
	\end{figure}

Our intention is to obtain the normalized solution of the modified Dirac equation below, $\psi$,  in the region between two parallel plates separated by a distance $a$. The plates are identical and have a lateral area equal to $L^2$, being $L>>a$, as shown in Fig. \ref{fg:plates}.

By applying \eqref{eq:lagrangian_violation} in Euler-Lagrange equation we get the equation that governs the dynamics of the fermionic field:
\begin{equation}
	\label{eq:EDP_all_cases}
	\left[ i\gamma^{0}\partial_t + i^{\xi} l^{\xi - 1} \left( \gamma^j \partial_j  \right)^{\xi} \right]\psi = 0.
	\end{equation}
	
In order to analyze the solutions of the above equation, we want first to point out that the structure for this equation becomes different for $\xi$ assuming even or odd values. For this reason, we will study each case separately. We will also require that all solutions satisfy the standard fermionic  MIT bag model boundary condition, given below:
\begin{equation}
\psi_{plates} = \left( i\gamma^{\mu}n_{\mu}\psi \right)_{plates} \ ,
\label{eq:MIT_bag}
\end{equation}
where the four-vector $n^{\mu}$ in the above equation always points to the orthogonal direction of the bag surface. For our case it reads:
\begin{equation}
n^\mu = \left\{ \begin{array}{ll}
(0,0,0,-1) & \quad \textrm{in the bottom plate} \left( z=0 \right), \\ 
(0,0,0,+1) & \quad \textrm{in the top plate} \left( z=a \right).
\end{array}  \right.
\end{equation}

In the case where there is no Lorentz violation, the MIT bag boundary condition on the surface prevents flux of fields cross this surface.\footnote{In \cite{DeTar:1979rg} there is a mathematical formulation of \eqref{eq:MIT_bag} by using the Dirac Lagrangian.} As we will see, this mathematical structure of confinement is also suitable for the fermionic fields described by \eqref{eq:EDP_all_cases}. All details that ensure the truthfulness of this remark will be discussed in Appendix \ref{app}. For this moment, the important thing to keep in mind is that Equation \ref{eq:MIT_bag} confines the field $\left( \psi \right)$ in the region between the two plates of Figure \ref{fg:plates}.

\subsection{Functional Form of the Fermionic Field for Even Values of $\xi$}
When $\xi$ is even number, the spacial differential part of \eqref{eq:EDP_all_cases} can be rewritten as follows:
	\begin{equation}
	\left( \gamma^j \partial_j \right)^{\xi}\psi = \left( \gamma^a \gamma^b \partial_a \partial_b \right)^{\xi /2} \psi =  \left( - \nabla^2 \right)^{\xi / 2}\psi.
	\end{equation}
Consequently, the Equation \eqref{eq:EDP_all_cases} becomes:
	\begin{equation}
	\label{eq:EDP_even_cases}
	\left[ i\gamma^0 \partial_t + i^{\xi} l^{\xi - 1} \left( - \nabla^2 \right)^{\xi /2} \right]\psi=0.
	\end{equation}
To solve the above equation, we adopt solutions with the following standard form:
	\begin{equation}
	\label{eq:solution_form}
	\psi \left(\textit{x} \right) =  
	\left(
	\begin{array}{c}
	\varphi(\textbf{x}) \\ 
	\chi(\textbf{x})
	\end{array}
	\right)e^{-iEt}.
	\end{equation}
For this case Eq. \eqref{eq:EDP_even_cases} becomes:
	\begin{eqnarray}
	E \varphi + l^{\xi -1} \left( \nabla^2 \right)^{\xi / 2} \varphi = 0, \\ 
	E \chi - l^{\xi -1} \left( \nabla^2 \right)^{\xi / 2} \chi = 0.
	\end{eqnarray}
	
Without an associated potential, the two-component spinors, $\varphi$ and $\chi$, must have a free waves functional form, i.e.,  $\nabla^2 \varphi = - \textbf{k}^2 \varphi$ and $\nabla^2 \chi = -\textbf{k}^2 \chi$. Using this feature we can rewrite the above equations in the following form:
	\begin{eqnarray}
	\left[ E + l^{\xi -1} \left( - \textbf{k}^2 \right)^{\xi / 2} \right] \varphi = 0, \\ 
	\left[ E - l^{\xi -1} \left( - \textbf{k}^2 \right)^{\xi / 2} \right] \chi = 0.
	\label{eq:system_even_case}
	\end{eqnarray}
	
In our analysis we will consider positive-energy solution, $ E \equiv \omega > 0$, and negative-energy  solutions, $ E \equiv - \omega < 0$, separately,  with corresponding two-component spinor, $\varphi^{(+)}$ and $\chi^{(+)}$, and $\varphi^{(-)}$ and $\chi^{(-)}$, respectively. First, we will study in details the solutions with positive energy, $\psi^{(+)}$, when $\xi / 2 = 1, 3, 5, \cdots$. In this case, the above equations become:
	\begin{eqnarray}
	\left[ \omega - l^{\xi -1} \left( \textbf{k}^2 \right)^{\xi / 2} \right] \varphi^{(+)} = 0, \\ 
	\left[ \omega + l^{\xi -1} \left( \textbf{k}^2 \right)^{\xi / 2} \right] \chi^{(+)} = 0.
	\end{eqnarray}
	
Assuming that $l$ is a positive parameter,\footnote{We could also assume $l$ being negative; although, it is not necessary. This choice does not create a new set of solutions to the problem in question.} we conclude that this system has no-trivial solutions if:
	\begin{eqnarray}
	\chi^{(+)}=0 
	\end{eqnarray}
	and
	\begin{eqnarray}
	\label{eq:energia}
	\omega= l^{\xi - 1}\left( \textbf{k}^2 \right)^{\xi / 2} \ .
	\end{eqnarray}
	
The boundary condition \eqref{eq:MIT_bag} requires that $\varphi^{(+)}(x,y,0) = \varphi^{(+)}(x,y,a) = 0$. So, normalized solutions that obey all desired criteria are expressed as: 
	\begin{eqnarray}
	\varphi^{(+)}_{r,\textbf{k}}&=& \frac{1}{\pi \sqrt{2a}} A_r e^{i(k_x x + k_y y)} \sin \left( k_z z \right)\nonumber\\
	& =& \frac{1}{\pi \sqrt{2a}} A_r e^{i(k_x x + k_y y)} \sin \left( \frac{n \pi z}{a}  \right) \ , \ n=1, \ 2, \ 3 \ \cdots.
	\label{eq:phi_plus_xi_2_6_10}
	\end{eqnarray}
	
The label $r$ in above equation indicates that for each momentum, $\textbf{k}$, there exist two independent solutions: $\varphi^{(+)}_{1,\textbf{k}}$ with $A_1 \equiv \textrm{col} \left( 1,0 \right)$ and $\varphi^{(+)}_{2,\textbf{k}}$ with $A_2 \equiv \textrm{col} \left( 0,1 \right)$. 
	
It is worth to mention that by using Eq. \eqref{eq:system_even_case} and boundary condition \eqref{eq:MIT_bag}, it is possible to find the fermionic field for both cases of energy. So, we have first to define what kind of solution we want to analyze, if positive- or negative-energy solution. This information in \eqref{eq:system_even_case} will eliminate one of the two-spinors: $\varphi^{(-)} = 0$ when $\xi = 2, 6, 10, \cdots$ and $\varphi^{(+)} = \chi^{(-)} = 0$ when $\xi = 4, 8,12 \cdots$. In all this cases the dispersion relation is $\omega = l^{\xi - 1} (\textbf{k}^2)^{\xi /2}$. The next step is to assume that the remaining two-component spinor has the free wave functional form, i.e.,
\begin{gather}
\chi^{(-)} = e^{-i (k_x x + k_y y)} ( A^{(-)}e^{ik_zz} + B^{(-)}e^{-ik_zz}) \quad \textrm{if} \quad \xi = 2,6,10, \cdots,\\
\chi^{(+)} = e^{i (k_x x + k_y y)} ( A^{(+)}e^{ik_zz} + B^{(+)}e^{-ik_zz}) \quad \textrm{if} \quad \xi = 4,8,12, \cdots,\\
\varphi^{(-)} = e^{-i (k_x x + k_y y)} ( A^{(-)}e^{ik_zz} + B^{(-)}e^{-ik_zz}) \quad \textrm{if} \quad \xi = 4,8,12, \cdots.
\end{gather}

 Now, it is necessary to use the boundary condition \eqref{eq:MIT_bag} to obtain discretized  value for $k_z$, and to find a relation between the two-component spinors $A^{(\pm)}$ and $B^{(\pm)}$. In all cases we found,
 \begin{eqnarray}
 k_z = n\pi /a \ {\rm and} \  B^{(\pm)} = - A^{(\pm)} \  .
 \end{eqnarray}
 The last step is to normalize the fermionic field. The normalization constant is $1/(\pi \sqrt{2a})$ and the matrix $A^{(\pm)}$ defines the polarization of the solution.
		
 The functional form of the wave function $\psi^{(\pm)}_{r,\textbf{k}}$ for any even value of $\xi$ is given bellow:
	\begin{equation}
	\psi^{(+)}_{r,k_x,k_y,n} = u_r  \frac{e^{i(k_x x + k_y y)}}{\pi \sqrt{2a}} \sin \left( \frac{n\pi z}{a} \right) e^{-i \omega_{\textbf{k}}t}.
	\label{eq:psi_plus_generic}
	\end{equation}
	\begin{equation}
	\psi^{(-)}_{r,k_x,k_y,n} = v_r \frac{e^{-i(k_x x + k_y y)}}{\pi \sqrt{2a}} \sin \left( \frac{n\pi z}{a} \right) e^{i \omega_{\textbf{k}}t}.
	\label{eq:psi_negativo_geral}
	\end{equation}
	
These solutions are already normalized. The four-component spinors, $u_r$ and $v_r$, are associated with the upper and lower components. Their explicit structures depend on the value of $\xi$. They are summarized in Table \ref{tab:u_r_and_v_r_even_cases}.
	\begin{table*}[h!]
	\centering
	\begin{tabular}{|c|c|c|c|c|}
	\hline 
	$\xi$ & $u_1$ & $u_2$ & $v_1$ & $v_2$ \\ 
	\hline 
	$2,6,10, \cdots$ & $ \left(\begin{array}{c}
	1 \\ 
	0 \\ 
	0 \\ 
	0
	\end{array} \right)  $ & $ \left( \begin{array}{c}
	0 \\ 
	1 \\ 
	0 \\ 
	0
	\end{array} \right)  $ & $ \left( \begin{array}{c}
	0 \\ 
	0 \\ 
	1 \\ 
	0
	\end{array} \right)  $ & $ \left( \begin{array}{c}
	0 \\ 
	0 \\ 
	0 \\ 
	1
	\end{array} \right) $ \\ 
	\hline 
	$4,8,12, \cdots$ & $ \left( \begin{array}{c}
	0 \\ 
	0 \\ 
	1 \\ 
	0
	\end{array} \right) $ & $ \left( \begin{array}{c}
	0 \\ 
	0 \\ 
	0 \\ 
	1
	\end{array} \right)  $ & $ \left( \begin{array}{c}
	1 \\ 
	0 \\ 
	0 \\ 
	0
	\end{array} \right) $ & $ \left( \begin{array}{c}
	0 \\ 
	1 \\ 
	0 \\ 
	0
	\end{array} \right) $ \\ 
	\hline 
	\end{tabular}
	\caption{The explicit forms for the four-component spinors $u_r$ e $v_r$ for specific even values of $\xi$.}
	\label{tab:u_r_and_v_r_even_cases}
	\end{table*}
\subsection{Functional Form of the Fermionic Field for Odd Values of $\xi$}

In this subsection, we will assume that $\xi$ is an odd number. In this case, the spatial differential part of \eqref{eq:EDP_all_cases} can be rewritten as follows:
\begin{equation}
	\left( \gamma^j \partial_j \right)^{\xi} = \left[\left( \gamma^j \partial_j \right)^2 \right]^{\frac{\xi -1}{2}} \gamma^i \partial_i = \gamma^j \partial_j \left( - \nabla^2  \right)^{\frac{\xi -1}{2}}.
	\end{equation}
	
Then, we can rewrite Eq. \eqref{eq:EDP_all_cases} in the following form:
	\begin{equation}
	\left[ i\gamma^0 \partial_t + i l^{\xi -1} \gamma^j \partial_j \left( \nabla^2 \right)^{\frac{\xi -1}{2}} \right] \psi = 0.
	\label{eq:EDP_odd_case}
	\end{equation}
	
To solve this equation, we will continue to adopt solutions in the form \eqref{eq:solution_form}. We begin investigating the solutions with positive energy $\left( E = \omega >0 \right)$:
	\begin{equation}
	\omega \varphi^{(+)} + i l^{\xi - 1} \sigma^j \partial_j \left( \nabla^2 \right)^{\frac{\xi -1}{2}} \chi^{(+)}=0 \   ,
	\label{Eq.1}
	\end{equation}
	\begin{equation}
	\omega \chi^{(+)} + i l^{\xi - 1} \sigma^j \partial_j \left( \nabla^2 \right)^{\frac{\xi -1}{2}}\varphi^{(+)} =0 \  . 
	\label{Eq.2}
	\end{equation}

Substituting $\chi^{(+)}$ given \eqref{Eq.2} into \eqref{Eq.1}, we find the following differential equation:
	\begin{equation}
	\omega^2 \varphi^{(+)} + l^{2(\xi -1)} \left( \nabla^2 \right)^{\xi} \varphi^{(+)} = 0.
	\label{eq:EDP_phi_plus}
	\end{equation}
	
We will adopt solutions in the following functional form:
	\begin{equation}
	\varphi^{(+)} = e^{i(k_x x + k_y y)} \left( A^{(+)}e^{ik_z z} + B^{(+)}e^{-ik_z z} \right).
	\label{eq:phi_plus_solution}
	\end{equation}
Where $A^{(+)}$ and $B^{(+)}$ are two-component spinors that will be determined later.
	
With \eqref{eq:EDP_phi_plus} and \eqref{eq:phi_plus_solution} we determine the dispersion relation: $\omega = l^{\xi - 1} \left( \textbf{k}^2 \right)^{\xi /2}$. The structure of the dispersion relation is the same for the even or odd case.
	
Now, substituting \eqref{eq:phi_plus_solution} into \eqref{Eq.2} we find the functional form for  $\chi^{(+)}$:
	\begin{equation}
	\chi^{(+)} = \frac{l^{\xi - 1}\left( -\textbf{k}^2 \right)^{\frac{\xi - 1}{2}}}{\omega} e^{i(k_x x + k_y y)} \left[ Q(k_z)A^{(+)}e^{ik_z z} + Q(-k_z)B^{(+)}e^{-ik_z z} \right],
	\end{equation}
	\noindent
where we use the notation,
\begin{eqnarray}
Q(\pm k_z) \equiv \sigma^1 k_x + \sigma^2 k_y \pm \sigma^3 k_z \  .
\end{eqnarray}
	
In the plate $ z=0 $ the boundary condition \eqref{eq:MIT_bag} relates $B^{(+)}$ to  $A^{(+)}$ in the following form:
	\begin{equation}
	B^{(+)} = -M_2^{-1} M_1 A^{(+)},
	\label{eq:_B_plus}
	\end{equation}
	\noindent
where $M_1$ and $M_2$ are $2 \times 2$ matrices defined bellow.   
	\begin{equation}
	\begin{split}
	&M_1 \equiv \omega I -il^{\xi - 1} \left( -\textbf{k}^2 \right)^{\frac{\xi - 1}{2}} \sigma^3 Q(k_z);\\
	&M_2 \equiv \omega I -il^{\xi - 1} \left( -\textbf{k}^2 \right)^{\frac{\xi - 1}{2}} \sigma^3 Q(-k_z);\\
	&M_3 \equiv \omega I +il^{\xi - 1} \left( -\textbf{k}^2 \right)^{\frac{\xi - 1}{2}} \sigma^3 Q(k_z);\\
	&M_4 \equiv \omega I +il^{\xi - 1} \left( -\textbf{k}^2 \right)^{\frac{\xi - 1}{2}} \sigma^3 Q(-k_z)  \  .
	\end{split}
	\label{eq:M_matrices}
	\end{equation}
Also, we have defined the $2\times 2$ matrices $M_3$ and $M_4$ that we will need later.
	
In the plate $z=a$ the boundary condition \eqref{eq:MIT_bag} provides another relation between  $A^{(+)}$ and $B^{(+)}$. It is
\begin{equation}
	M_3 A^{(+)} e^{ik_z a} + M_4 B^{(+)} e^{-ik_z a} = 0  \  .
	\end{equation}
	
This new equation together with Eq. \eqref{eq:_B_plus} determine the discretization of $k_z$:
\begin{eqnarray}
	&&\left[ Ie^{ik_z a} - M_3^{-1} M_4 M_2^{-1} M_1 e^{-ik_z a} \right] A^{(+)} =0\ , \nonumber \\
	 &&\left( e^{ik_z a} + e^{-ik_z a} \right)A^{(+)} = 0 \  \Rightarrow
	k_z = \left( n + \frac{1}{2} \right)\frac{\pi}{a}\cdot
	\label{eq:quantization_k_z_odd}
	\end{eqnarray}
In the above expressions, $n$ can be zero or any positive integer number.
	
There are two linearly independent solutions for each $\textbf{k}$, i.e., $\psi^{(+)}_{1,\textbf{k}}$ and $\psi^{(+)}_{2,\textbf{k}}$. Consequently, $\varphi^{(+)}$ and $\chi^{(+)}$  must have an index $r $, that specifies the solution we are considering. These functions are exhibited below:
	\begin{equation}
	\varphi^{(+)}_{r,\textbf{k}} = \frac{e^{i(k_x x + k_y y)}}{4\pi \sqrt{a}}  \left[ A^{(+)}_r e^{ik_z z} + B^{(+)}_{r,\textbf{k}} e^{-ik_z z}\right]
	\end{equation}
and
	\begin{equation}
	\chi^{(+)}_{r,\textbf{k}} = \frac{l^{\xi - 1} \left( - \textbf{k}^2 \right)^{\frac{\xi - 1}{2}}}{\left(4 \pi \sqrt{a}\right) \omega_{\textbf{k}}} e^{i(k_x x + k_y y)} \left[ Q(k_z)A^{(+)}_r e^{ik_z z} + Q(-k_z)B^{(+)}_{r, \textbf{k}}e^{-ik_z z} \right]  \ .
	\end{equation}
The factor $\frac{1}{4\pi \sqrt{a}}$ comes from normalization, and $A^{(+)}_r$ and $B^{(+)}_{r,\textbf{k}}$ are two-component spinors. Their structures are provided in the first column of Table \ref{tab:A_and_B_odd_case}.

Our next step is to obtain the negative-energy wave function. For this case, Eq. \eqref{eq:EDP_odd_case} becomes:
	\begin{eqnarray}
	\omega \varphi^{(-)} - i l^{\xi - 1} \sigma^j \partial_j \left( \nabla^2 \right)^{\frac{\xi - 1}{2}} \chi^{(-)} = 0  \ 
	\label{Eq.3}
	\end{eqnarray}
	and 
	\begin{equation}
	\omega \chi^{(-)} - i l^{\xi - 1} \sigma^j \partial_j \left( \nabla^2 \right)^{\frac{\xi - 1}{2}} \varphi^{(-)} = 0  \  .
	\label{Eq.4}
	\end{equation}
Substituting $\varphi^{(-)}$ from \eqref{Eq.3} into \eqref{Eq.4}, we get a  differential equation that determines $\chi^{(-)}$:
	\begin{equation}
	\omega^2 \chi^{(-)} + l^{2(\xi - 1)} \left( \nabla^2 \right)^\xi \chi^{(-)} = 0.
	\end{equation}
	
The above equation has the same structure of \eqref{eq:EDP_phi_plus}. Then, by analogy, the normalized negative-energy solution, $\psi^{(-)}$, can be given by the two-component spinors below:
	\begin{eqnarray}
	\varphi^{(-)}_{r,\textbf{k}} &=& \frac{l^{\xi - 1} \left( - \textbf{k}^2 \right)^{\frac{\xi - 1}{2}}}{\left(4 \pi \sqrt{a}\right) \omega_{\textbf{k}}} e^{-i(k_x x + k_y y) } \left[ Q(-k_z)A^{(-)}_r e^{ik_z z} + Q(k_z)B^{(-)}_{r, \textbf{k}}e^{-ik_z z} \right] \  , \nonumber\\
	\chi^{(-)}_{r,\textbf{k}} &=& \frac{e^{-i(k_x x + k_y y)}}{4\pi \sqrt{a}}  \left[ A^{(-)}_r e^{ik_z z} + B^{(-)}_{r,\textbf{k}} e^{-ik_z z}\right].
	\end{eqnarray}
Where the explicit forms of $A^{(-)}_r$ and $B^{(-)}_{r,\textbf{k}}$ are in the second column of Table \ref{tab:A_and_B_odd_case}.
	\begin{table*}[h!]
	\centering
	\begin{tabular}{|c|c|c|}
	\hline 
	 & $E = \omega$ & $E =- \omega$ \\ 
	\hline 
	$A_1$ & $\left( \begin{array}{c}
	1 \\ 
	0
	\end{array}  \right)$ & $\left( \begin{array}{c}
	1 \\ 
	0
	\end{array}  \right)$ \\ 
	\hline 
	$A_2$ & $\left( \begin{array}{c} 
	0 \\ 
	1
	\end{array} \right)$ & $\left( \begin{array}{c} 
	0 \\ 
	1
	\end{array} \right)$ \\ 
	\hline 
	$B_{1,\textbf{k}}$ & $\frac{l^{\xi -1} \left( - \textbf{k}^2 \right)^{\frac{\xi - 1}{2}}}{i \omega_{\textbf{k}}} \left( \begin{array}{c}
k_z \\ 
-(k_x + ik_y)
\end{array}  \right)$ & $- \frac{l^{\xi -1} \left( - \textbf{k}^2 \right)^{\frac{\xi - 1}{2}}}{i \omega_{\textbf{k}}} \left( \begin{array}{c}
k_z \\ 
k_x + ik_y
\end{array}  \right)$ \\ 
	\hline 
	$B_{2,\textbf{k}}$ & $\frac{l^{\xi -1} \left( - \textbf{k}^2 \right)^{\frac{\xi - 1}{2}}}{i \omega_{\textbf{k}}} \left( \begin{array}{c}
k_x - ik_y \\ 
k_z
\end{array}  \right)$ & $- \frac{l^{\xi -1} \left( - \textbf{k}^2 \right)^{\frac{\xi - 1}{2}}}{i \omega_{\textbf{k}}} \left( \begin{array}{c}
-(k_x - ik_y) \\ 
k_z
\end{array}  \right)$ \\ 
	\hline 
	\end{tabular}
	\caption{The set of two-component spinors that appear in the expressions of fermionic fields confined between two parallel plates, which obey the MIT bag boundary condition on the two plates and Lorentz symmetry violation in the manner proposed by Horava-Lifshitz.}
	\label{tab:A_and_B_odd_case}
	\end{table*}
\section{Casimir Energy}
	\label{sec:Casimir_effect}
The objective of this section is to obtain the fermionic Casimir energy and pressure. Adopting the formalism of \cite{Itzykson:1980rh}, we write below the explicit expression for the field operator:
	\begin{equation}
	\psi (x) = \sum_{r, n} \int dk_x \int dk_y \left[ c_r(\textbf{k}) \psi^{(+)}_{r, \textbf{k}} + d_r^{\dag}(\textbf{k})\psi^{(-)}_{r, \textbf{k}}\right] \   ,
	\label{field_operator}
	\end{equation}
where  $c_r(\textbf{k})$ is the operator that destroys particle with momentum $\textbf{k}$ and polarization $r$, while $d^{\dag}_r(\textbf{k})$ creates antiparticle with momentum $\textbf{k}$ and polarization $r$.
	
It has been shown in \cite{Mandl:1985bg} that the standard Dirac Lagrangian provides the following Hamiltonian density operator:
	\begin{equation}
	\mathcal{H} = i \psi^{\dag} \dot{\psi}.
	\end{equation}
By adopting a similar procedure, we can prove that the modified Lagrangian \eqref{eq:lagrangian_violation}, provides also the same expression for the Hamiltonian density.
	
The vacuum energy, $E_0$, is given by taking the expected value of the Hamiltonian operator in the vacuum state:
	\begin{equation}
	E_0 = \left< 0 \left| H \right| 0 \right> = i \int_V d^3x \left< 0 \left| \psi^{\dag} \dot{\psi} \right| 0 \right>.
	\end{equation}
	
The MIT bag boundary condition confines the field in the region between the two parallel plates. Because we are considering large plates, we can approach the volume of integration by $V=aL^2$. Since the Casimir effect is detectable only when the separation of plates is very small, we have to assume that $L>>a$.
Substituting the field operator \eqref{field_operator} into the above equation, and after some intermediate steps, we obtain:
\begin{equation}
	E_0 =  \sum_{r} \int_{- \infty}^{\infty} dk_x \int_{- \infty}^{\infty} dk_y \sum_{n} \omega_{\textbf{k}} \left< 0 \left| c^{\dag}_r(\textbf{k}) c_r(\textbf{k}) - d_r(\textbf{k}) d^{\dag}_r(\textbf{k}) \right| 0 \right> \cdot
	\label{Vacuum_energy}
	\end{equation}
In the above calculation we use the following algebra for the fermionic creation and annihilation operators: 
\begin{eqnarray}
\label{algebra}
\{c_{\textbf{k},n}, c^{\dagger}_{\textbf{q},m}\} & = \delta_{n,m}\delta^2(\textbf{k}-\textbf{q}) , \nonumber \\
\{d_{\textbf{k},n}, d^{\dagger}_{\textbf{q},m}\} & = \delta_{n,m}\delta^2(\textbf{k}-\textbf{q})  \  ,
\end{eqnarray}
with all other anti-commutation relations being zero.	
	
Using Eq. \eqref{algebra} we rewrites \eqref{Vacuum_energy} as follows:
	\begin{equation}
	E_0 = - \frac{L^2  l^{\xi - 1}}{4 \pi^2} \sum_{r, n} \int_{- \infty}^{\infty} dk_x \int_{- \infty}^{\infty} dk_y \left( \textbf{k}^2 \right)^{\xi / 2}.
	\end{equation}
	
By expressing $k_x$ and $k_y$ in polar coordinates the previous equation becomes:
	\begin{equation}
	E_0 = - \frac{L^2}{\pi} l^{\xi - 1} \int_0^{\infty} dk \ k\sum_n \left( k^2 + k_z^2 \right)^{\xi / 2} \  .
	\label{Vacuum_energy1}
	\end{equation}
	
As we have already mentioned the discrete values assumed by $k_z$ depends if the critical exponent, $\xi$, is an even or odd number. For this reason, we will calculate \eqref{Vacuum_energy1}, separately, for each case.
\subsection{Casimir Energy when $\xi$ is Even}

Considering an even value for $\xi$, the momentum along the $z-$direction is discretized according to $k_z=n\pi/a$, for $n=1, \ 2, \ 3, \cdots$ .
The procedure adopted in this paper to calculate the summation over the quantum number $n$ in \eqref{Vacuum_energy1} is through the Abel-Plana summation formula given below \cite{Bordag:2009zzd}: 
 \begin{equation}
 \sum_{n=1}^{\infty} f(n) = -\frac{1}{2} f(0) + \int_0^{\infty} dt f(t) + i \int_0^{\infty} \frac{f(it) - f(-it)}{e^{2\pi t} - 1} dt\cdot
 \label{eq:abel_plana_n}
 \end{equation}
	
For our case,
	\begin{equation}
	f(n) = \left[ k^2 + \left( \frac{n\pi}{a} \right)^2 \right]^{\xi/2} \ . 
	\end{equation}
Consequently,
	\begin{equation}
	\label{AP}
	\sum_{n=1}^{\infty} f(n) = - \frac{1}{2} k^{\xi} + \int_0^{\infty} dt \left[ k^2 + \left( \frac{t\pi}{a} \right)^2 \right]^\frac{\xi}{2} + i \int_0^{\infty} \frac{[k^2 + (it\pi /a)^2]^\frac{\xi}{2} - [k^2 + (-it\pi / a)^2]^\frac{\xi}{2}}{e^{2\pi t} - 1} dt.
	\end{equation}
	
The second term in the right-hand side of the above equation, when substituted in \eqref{Vacuum_energy1} provides a divergent result. This infinity energy corresponds to the vacuum energy without plates. Moreover, the first term also provides a divergent result. It corresponds to the vacuum energy in the presence of a single plate. Both terms do not contribute to the Casimir energy. So, we discard them. The Casimir energy, $E_c$, is $E_0$ renormalized. This means that all divergent terms in $E_0$ must be subtracted. So the Casimir energy corresponds the energy due to the two plates, and is given by:
	\begin{equation}
	E_c = -\frac{iL^2l^{\xi - 1}}{\pi} \int_0^{\infty} dk k  \int_0^{\infty} \frac{[k^2 + (it\pi /a)^2]^{\xi / 2} - [k^2 + (-it\pi / a)^2]^{\xi / 2}}{e^{2\pi t} - 1} dt  \  .
	\label{Energy_C0}
	\end{equation}
	
In order to evaluate the Casimir energy above, we will use Euler's formula \cite{hassani2008mathematical} to compute the expression below:
	\begin{equation}
	\left[ k^2 + (\pm i u)^2 \right]^\frac{\xi}{2} = \left\{ \begin{array}{cl}
	( k^2 - u^2 )^{\xi / 2} \ ,  & \textrm{if $u<k$ } \ . \\ 
	e^{\pm i \pi \xi /2}(u^2 - k^2)^{\xi / 2} \ ,  & \textrm{if $u>k$} \  .
	\end{array}   \right.
	\label{Euler}
	\end{equation}
	
The immediate implication of \eqref{Euler} is that both terms in the integrand of \eqref{Energy_C0} cancel, consequently, the  Casimir energy vanishes for all values of even $\xi$. This fact also happens for massless scalar field as shown in \cite{Ferrari:2010dj,Ulion:2015kjx}. 
\subsection{Casimir Energy when $\xi$ is Odd}
When $\xi$ assumes odd values, we have to assume $k_z=\left(n + \frac{1}{2}\right) \frac{\pi}{a}$ in the equation \eqref{Vacuum_energy1}. In this case to develop the summation over the quantum number $n$ we use the Abel-Plana summation formula below \cite{Bordag:2009zzd}:
	\begin{equation}
	\sum_{n=0}^{\infty}  f(n+1/2) = \int_0^{\infty} dt f(t) -i \int_0^{\infty} dt \frac{f(it) - f(-it)}{e^{2\pi t} + 1} \cdot
	\label{eq:abel_plana_half_integer}
	\end{equation}
	
For this case,
\begin{equation}
	f(n+1/2) = \left\{ k^2 + \left[ \left( n + \frac{1}{2} \right)\frac{\pi}{a} \right]^2 \right\}^{\xi/2}.
	\end{equation}
	
Consequently,
	\begin{equation}
	\sum_{n=0}^{\infty}f(n+1/2) = \int_0^{\infty} dt \left[ k^2 + (t \pi / a)^2 \right]^{\xi /2} - i \int_0^{\infty} \frac{\left[ k^2 + \left( \frac{it \pi}{a} \right)^2 \right]^\frac{\xi}{2} - \left[ k^2 + \left( \frac{-it \pi}{a} \right)^2 \right]^\frac{\xi}{2}}{e^{2\pi t}+1}dt.
	\end{equation}
	
There appears a divergent contribution in $E_0$ when the first integral in the right-hand side is inserted into \eqref{Vacuum_energy1}. The renormalized Casimir energy is obtained by subtracting from $E_0$ this divergence. Then, the Casimir energy is given only by the second integral. For mathematical convenience we will make the following change of variable, $u = \frac{\pi t}{a}$. So, we obtain:
	\begin{equation}
	E_c = \frac{iaL^2l^{\xi - 1}}{\pi^2} \int_0^{\infty} dk k \int_0^{\infty} du \frac{[k^2+(iu)^2]^{\xi / 2} - [k^2+(-iu)^2]^{\xi / 2}}{e^{2au}+1} \cdot
	\end{equation}
	
The integral over the variable $u$ can be divided in two sub-intervals: From $[0, \ k]$ and from $[k, \ \infty)$. According to \eqref{Euler}, the contribution from the first segment vanishes. So it remains the contribution from the second interval. As a consequence, the Casimir energy can be written as: 
	\begin{equation}
	E_c = - \frac{2aL^2l^{\xi - 1}}{\pi^2} \sin \left( \frac{\xi \pi}{2} \right) \int_0^{\infty} dk k  \int_k^{\infty}\frac{(u^2 - k^2)^{\xi / 2}}{e^{2au}+1}du.
	\end{equation}
	
By using the new variable $u \equiv kt$ the above expression can be written in a  more suitable form:
	\begin{equation}
	E_c = \frac{-2aL^2l^{\xi - 1}}{\pi^2} \sin \left( \frac{\xi \pi}{2} \right) \int_1^{\infty} dt (t^2 - 1)^{\xi / 2} \int_0^{\infty} dk \frac{k^{\xi + 2}}{e^{2akt}+ 1}\cdot
	\end{equation}
	
To calculate this expression, we will use the following formula taken from \cite{jeffrey2007table}:
	\begin{equation}
	\int_0^{\infty} \frac{t^{z-1}}{e^t + 1} = \left( 1 - 2^{1-z} \right) \Gamma(z)\zeta(z) \ , \quad  \textrm{for $Re (z) > 0$} \ ,
	\end{equation}
where $\Gamma(z)$ and $\zeta(z)$, represent the Gamma and Riemann Zeta functions, respectively.
	
Consequently,
	\begin{equation}
	E_c = - \frac{L^2l^{\xi - 1}}{\pi^2 (2a)^{\xi + 2}} \left[ 1 - \frac{1}{2^{\xi + 2}}  \right] \sin \left( \frac{\xi \pi}{2} \right) \Gamma (\xi + 3) \zeta (\xi + 3) \int_1^{\infty}dt \frac{(t^2 - 1)^{\xi / 2}}{t^{\xi +3}} \cdot
	\end{equation}
The remaining integral has a closed solution:
	\begin{equation}
	\int_1^{\infty} dt  \frac{(t^2 - 1)^{\xi / 2}}{t^{\xi +3}} = \left. \frac{(t^2-1)^{\frac{\xi +2}{2}}}{(\xi + 2)t^{\xi + 2}} \right|_1^{\infty} = \frac{1}{\xi + 2} \  .
	\end{equation}
Finally, we get:
	\begin{equation}
	E_c = - \sin \left( \frac{\xi \pi}{2} \right) \frac{L^2l^{\xi - 1}}{\pi^2 (2a)^{\xi+2}} \left( 1 - \frac{1}{2^{\xi + 2}} \right)\Gamma (\xi + 2) \zeta (\xi + 3).
	\label{eq:Casimir_energy_all_cases}
	\end{equation}
Although we have obtained this expression assuming that $\xi$ is an odd number, it is also valid when $\xi$ is even.
	
When $\xi = 1$, Eq. \eqref{eq:Casimir_energy_all_cases} reproduces the well known result for the fermionic  Casimir energy per unit area associated with massless field \cite{DePaola:1999im}:
	\begin{eqnarray}
	\frac{E_c}{L^2}  = - \frac{7}{4} \left( \frac{\pi^2}{720a^3} \right) \ .
	\end{eqnarray}
	
The Casimir pressure is:
\begin{equation}
	P_c = - \frac1{L^2}\frac{\partial E_c}{\partial a} = - \sin \left( \frac{\xi \pi}{2} \right) \frac{2l^{\xi - 1}}{\pi^2 (2a)^{\xi+3}} \left( 1 - \frac{1}{2^{\xi + 2}} \right) \Gamma (\xi + 3) \zeta (\xi + 3) \ .
	\label{eq:casimir_force}
\end{equation}
It is worth to  mention that the resulting expression above is an analytic functions of $\xi$ and by analytic continuation is valid for all values of $\xi$. Consequently the null result obtained for even $\xi$ is contained in this expression. Moreover, due to the sin function, for the value of $\xi=1, \ 5, \ 9, \ ...$, the force is attractive, and for $\xi=3, \ 7, \ 11, \ ... $, the Casimir force is repulsive. 
	
\begin{figure}[hbtp]
	\centering
	\includegraphics[scale=0.8]{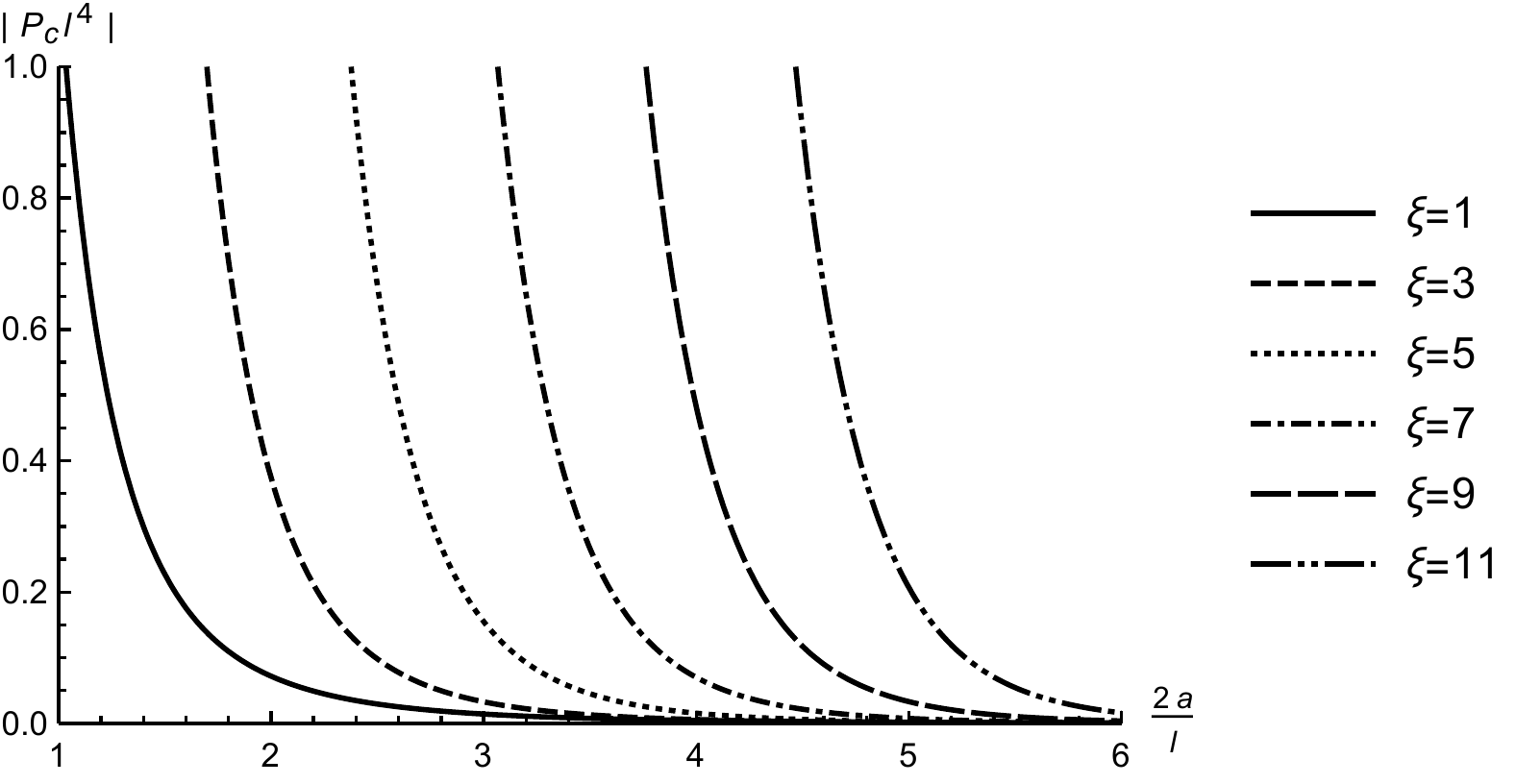}
	\caption{The Casimir pressure $|P_c|l^4$  as function of $2a/l$ for different values of $\xi$. }
	\label{Pressure}
\end{figure}

The Casimir pressure is detectable only when the distance between the two plates is very small. Equation \eqref{eq:casimir_force} shows us that the critical exponent, $\xi$, increases the intensity of it.\footnote{By direct numerical calculation we can see that $\frac{(|P_c|l^4)_{\xi=3}}{(|P_c|l^4)_{\xi=1}}\approx 10^3$ for $2a/l=0.1$.} In order to exhibit the intensity of the Casimir pressure with the distance between the two plates, for different values of critical exponent, we plot in Fig. \ref{Pressure} the modulus of the Casimir pressure in units of $1/l^4$ as function of the dimensionless parameter $2a/l$ for different values of $\xi$.

The expression for the Casimir pressure \eqref{eq:casimir_force}, is similar to the corresponding one obtained for a massless scalar field in an H-L approach, for the case where the scalar quantum field obeys the Dirichlet boundary condition in one plate, and Newman boundary condition in the other \cite{Ulion:2015kjx}.  The only difference is a $-\frac{1}{4}$ factor. In fact this similarity resides in the discretization expression for the component of the momentum perpendicular to the plates. For both cases we have found $k_z=(n+1/2)\pi/a$. However, both fields operator, bosonic and fermionic, obey different statistics, and also different number degree of freedom, resulting in a different global factor  multiplying the Casimir pressure. 
	
\section{Conclusions and Remarks}
	\label{sec:conclusions}
	
The Casimir forces associated with electromagnetic, scalar and fermionic fields in a standard quantum field theory, i.e.,  without Lorentz violation, is inversely proportional to the fourth power of the distance between the plates. Using an H-L approach, that introduces a Lorentz symmetry breaking due to the asymmetry between time and space, the Casimir force was investigated in this paper, considering massless fermionic fields governed by the Eq.(\ref{eq:lagrangian_violation}). We found that Casimir force depends inversely with $a^{\xi + 3}$, being $\xi\geq 1$ (the critical exponent). Consequently, it is stronger than the case with Lorentz symmetry, $\xi=1$. In the literature, we find Casimir force being attractive (usual case) or repulsive \cite{Jiang:2018ivv}. It depends on the construction of the experiment. The Equation \eqref{eq:casimir_force} tell us that the critical  exponent, $ \xi $, change the attractive or repulsive nature of Casimir force. Another direct conclusion of this result is that more accurate experiments related to the detection of Casimir force may reveal if the Lorentz symmetry is, in fact, a symmetry of Nature.

 Although our calculations have been developed for massless fields, it our interest to consider, in near future, the analysis of Casimir energy considering massive fields. In the standard Lorentz preserving symmetry case, the discretization relation obeyed by the momentum perpendicular to the plates is given by a transcendental equation. So we expect that considering the H-L approach, this discretization relation for the case with $\xi$ being an integer number should be much more complicated. 
	
In Section \ref{secII}, we have used the MIT bag boundary condition, Eq. \eqref{eq:MIT_bag},  to construct normalized  positive- and negative-energy  fermionic wave-function. As it is well know, in the standard Lorentz preserving symmetry formalism, this condition prevents the fermionic current to cross the boundaries. So, an important point of this paper is to show that this same condition also prevent the generalized fermionic current obtained in the Horava-Lifshitz theory for fermionic fields to cross the two parallel plates. This issue is addressed in Appendix \ref{app}. There we prove that this fact happens, i.e., the generalized fermionic current vanishes on the plates.

{\bf Acknowledgements.} 
This work was partially supported by Conselho Nacional de Desenvolvimento
Científico e Tecnológico (CNPq). E. R. Bezerra de Mello has been partially supported by the CNPq through the project No. 313137/2014-5. D. R. da Silva and  M. B. Cruz have been supported by Coordena\c{c}\~ao de Aperfeiçoamento de Pessoal de N\'{\i}vel Superior (CAPES).

\appendix
\section{Modified Fermionic Current}
\label{app}
In this appendix, we want to show that standard MIT boundary condition provides the cancellation of the modified fermionic currents on the plates. In order to do that, it is necessary first to find the expressions for the currents by using the modified Dirac equation \eqref{eq:EDP_all_cases}. We will use a  similar process as presented in \cite{Bjorken:1965zz}.

We begin studying the fermionic current in the even cases, $ \xi = 2m $. So, taking Eq. \eqref{eq:EDP_even_cases} we get the modified continuity equation given below for the fermionic current:
\begin{equation}
\partial_t \left( \psi^{\dag} \psi \right) + il^{2m - 1} \left\{ \left[ \left( \nabla^2 \right)^m \bar{\psi}\right]\psi - \bar{\psi}\left( \nabla^2 \right)^m \psi \right\} \equiv \partial_t \rho  + \mathbf{\nabla} \cdot \textbf{J}=0 \  ,
\label{eq:F_n}
\end{equation}
with the corresponding modified vector current density,
\begin{equation}
\textbf{J} =  il^{2m - 1} \sum_{i=1}^{m} \left\{ \left[ \bm{\nabla}  \left( \nabla^2 \right)^{i-1}\bar{\psi} \right] \left( \nabla^2 \right)^{m - i}\psi  - \left[ \left( \nabla^2 \right)^{i-1}\bar{\psi} \right]\bm{\nabla}  \left( \nabla^2 \right)^{m - i}\psi \right\}.
\label{Current_even}
\end{equation}

As we have shown, the functional form of the wave-functions when $\xi$ is an even number presents a dependence on the $z-$variable through $\sin (n\pi z /a)$. Of course this form will not be changed by application of the Laplacian operator; on the other hand, the projection of \eqref{Current_even} along the $z-$direction will contain partial derivative with respect to this variable. So all contributions in $\hat{z}\cdot\textbf{J}$ will present explicit dependence on the product $\cos(n\pi z/a)\sin (n\pi z /a)$, which vanishes on the plates. So we conclude that the standard MIT bag boundary condition prevents the modified current to cross the flat boundaries.
 
Now, let us turn our attention to the fermionic current in odd cases of critical exponent,  $\xi = 2m + 1 $. Doing the same procedure as before to obtain  \eqref{eq:F_n}, however using the modified Dirac Equation \eqref{eq:EDP_odd_case}, we get:
\begin{equation}
\partial_t \left( \psi^{\dag}\psi \right) + l^{2m} \left\{ \bar{\psi}\gamma^j \partial_j \left( \nabla^2 \right)^m \psi + \left[ \partial_j \left( \nabla^2 \right)^m \bar{\psi} \right]\gamma^j \psi \right\} \equiv \partial_t \rho + l^{2m}G_m = 0.
\label{eq:G_n}
\end{equation}

The above equation ensures the existence of a modified continuity equation if there exists a vector current density, $J^j$, that satisfies the following relation:
\begin{equation}
\partial_jJ^j = l^{2m}G_m.
\label{eq:G_m_and_current_relation}
\end{equation}
Fortunately, we were able to obtain this current. It reads,
\begin{equation}
J^j = \frac{1}{2} [\tilde{J}^j + (\tilde{J}^j)^\dag],
\end{equation}
with
\begin{equation}
\begin{split}
\tilde{J}^j =& l^{2m}\bar{\psi}\gamma^j \left( \nabla^2 \right)^m \psi + l^{2m}\sum_{i=1}^{m} \left[ \partial_k \left( \nabla^2 \right)^{i-1} \bar{\psi} \right] \gamma^k \partial^j \left( \nabla^2 \right)^{m-i} \psi + \\
& -  l^{2m} \sum_{i=1}^{m} \left[ \partial^j \partial_k \left( \nabla^2 \right)^{i-1} \bar{\psi} \right]\gamma^k  \left( \nabla^2 \right)^{m-i} \psi.
\end{split}
\label{eq:current_odd_case}
\end{equation}

As expected, the above equation reproduces the standard vector Dirac current when $\xi = 1$, $J^j = \bar{\psi}\gamma^j \psi$. In this case, the cancellation of the current on the plates is already established. Working with the general expression to find this result is much more complicated. Our intention is to show that density current vanishes on the plates. We have already guaranteed the existence of probability conservation, consequently
\begin{eqnarray}
\frac{d}{dt}\int d^3x \rho + \int d^3x \bm{\nabla}\cdot \textbf{J} = 0 \Rightarrow \oint \textbf{J} \cdot d\textbf{a} = - \frac{d}{dt}\int d^3x \psi^\dag \psi = 0  \  .
\end{eqnarray}
Then, the total current flow on the plates is zero. This implies two possibilities: the current is zero in each surface, or the current flow in one plate is opposite of the flow in the other plate. If the second possibility is the correct one, the probability density on plates is not zero. However, the MIT bag boundary condition ensures that $\rho = \psi^\dag \psi$ is zero on the bag (the plates), so the only plausible explanation is that the vector current density, in fact, vanishes on each surface.

To complete our prove, we need to show that the integral of $\rho$ over space is constant in time. Let us take the most general solution of the Dirac equation, which is give in terms of a combination of positive- and negative-energy modes:
\begin{equation}
\psi = \sum_r \int dk_x \int dk_y \sum_{k_z} \left[ c(r,\bm{k})\psi^{(+)}_{r,\bm{k}} + d(r,\bm{k})\psi^{(-)}_{r,\bm{k}} \right]  \ .
\label{geenral_solution}
\end{equation}
where
\begin{gather}
\int \left(\psi^{(\pm)}_{r,\bm{k}}\right)^\dag \psi^{(\pm)}_{s,\bm{q}} = \delta_{rs}\delta(k_x-q_x) \delta(k_y-q_y) \delta_{k_z q_z},\\
\int \left(\psi^{(\pm)}_{r,\bm{k}}\right)^\dag \psi^{(\mp)}_{s,\bm{q}} = 0 \ .
\end{gather}
So, by using \eqref{geenral_solution} and the above results we have, 
\begin{equation}
\int d^3 x \psi^\dag \psi = \sum_{r} \int dk_x \int dk_y \sum_{k_z} \left[ |c(r,\bm{k})|^2 + |d(r,\bm{k})|^2 \right] \equiv 1.
\end{equation}

\end{document}